\documentclass[prb,twocolumn,aps,showpacs,fixfloats]{revtex4}
\usepackage{graphicx}
\usepackage{bm}
\usepackage{amsmath,amssymb}
\usepackage{subfigure}
\usepackage{float}
\usepackage{latexsym}
\usepackage{color}
\usepackage{enumerate}
\usepackage{pdfpages}
\usepackage{tikz}
\usepackage{hyperref}
\usepackage{graphicx}
\usepackage{bm}
\usepackage{amssymb} 
\usepackage{amsmath}
\usepackage{subfigure}
\usepackage{relsize}

\begin{document}
\newcommand{\s}{\scriptscriptstyle}
\newcommand{\uu}{\uparrow \uparrow}
\newcommand{\ud}{\uparrow \downarrow}
\newcommand{\du}{\downarrow \uparrow}
\newcommand{\dd}{\downarrow \downarrow}
\newcommand{\ket}[1] { \left|{#1}\right> }
\newcommand{\bra}[1] { \left<{#1}\right| }
\newcommand{\bracket}[2] {\left< \left. {#1} \right| {#2} \right>}
\newcommand{\vc}[1] {\ensuremath {\bm {#1}}}
\newcommand{\tr}{\text{Tr}}
\newcommand{\Trans}{\ensuremath \Upsilon}
\newcommand{\Refl}{\ensuremath \mathcal{R}}

\title{Landau-Zener transition between two quantum dots coupled  by resonant tunneling}

\author{M. E. Raikh}

\affiliation{ Department of Physics and
Astronomy, University of Utah, Salt Lake City, UT 84112}

\begin{abstract}
We consider the transition of electron between two quantum dots in which the discrete levels are swept past each other with a constant velocity.  If a direct tunneling between the dot levels was allowed, an electron will be transferred  between the dots when the levels cross. This transfer is described in terms of the conventional Landau-Zener theory.  We assume that  direct tunneling between the dots is forbidden. Rather, the transfer is due to the resonant tunneling via a discrete impurity level separating the dots.  Then the description of the electron transfer reduces to a three-state (two dots plus impurity) Landau-Zener transition. Transition probability depends on the relative positions of the resonant level  and the energy at which the levels cross.     It also depends on the left-right asymmetry of  tunneling between the impurity and the left(right) dots. We calculate the transition probability in different limits of the horizontal (in space) and vertical (in energy) impurity level positions.

\end{abstract}
\maketitle

\section{Introduction}

For  several decades after it was introduced, the  
Landau-Zener physics\cite{Landau,Zener,Majorana,Stukelberg} was invoked in  relation to the atomic collisions, see e.g. Refs. \onlinecite{atomic0, atomic1}.  In the past two decades the relevance of the Landau-Zener-based description was demonstrated  for qubits   based on   superconducting circuits\cite{dot1,dot2}  and  quantum dot interferometers,\cite{dot3,dot4,dot5}  see also the reviews \onlinecite{review1,review2}.   

On the theory side, the studies of the transitions taking place upon drive-induced crossing of two levels have expanded to include multistate Landau-Zener transitions, which  take place when several levels cross upon application of  a linear drive. 
In the course of analysis of the multistate transitions a number of models which
allow an exact solution had been identified.\cite{Demkov1968,Carroll,Carroll1,Bow1,Bow2,Bow3,Bow5,Sinitsyn2004,Shytov,Sinitsyn1,Sinitsyn2,Volkov,Sinitsyn3} 

One of realizations of a two-state Landau-Zener model is a solid state qubit representing a double  quantum dot, where two dots are coupled by tunneling and are driven by external voltage. In this paper we consider the simplest realization of the three-state model within the resonant-tunneling setup. Namely, we assume that a direct tunneling between the dots is forbidden. Rather, the electron transfer between the dots  is due to the resonant tunneling via a discrete impurity level separating them.  While the corresponding three-state model is  not solvable, we apply the perturbative and semiclassical descriptions to calculate the tunneling probability between the driven dot levels.

\begin{figure}
\label{F}
	\includegraphics[scale=0.70]{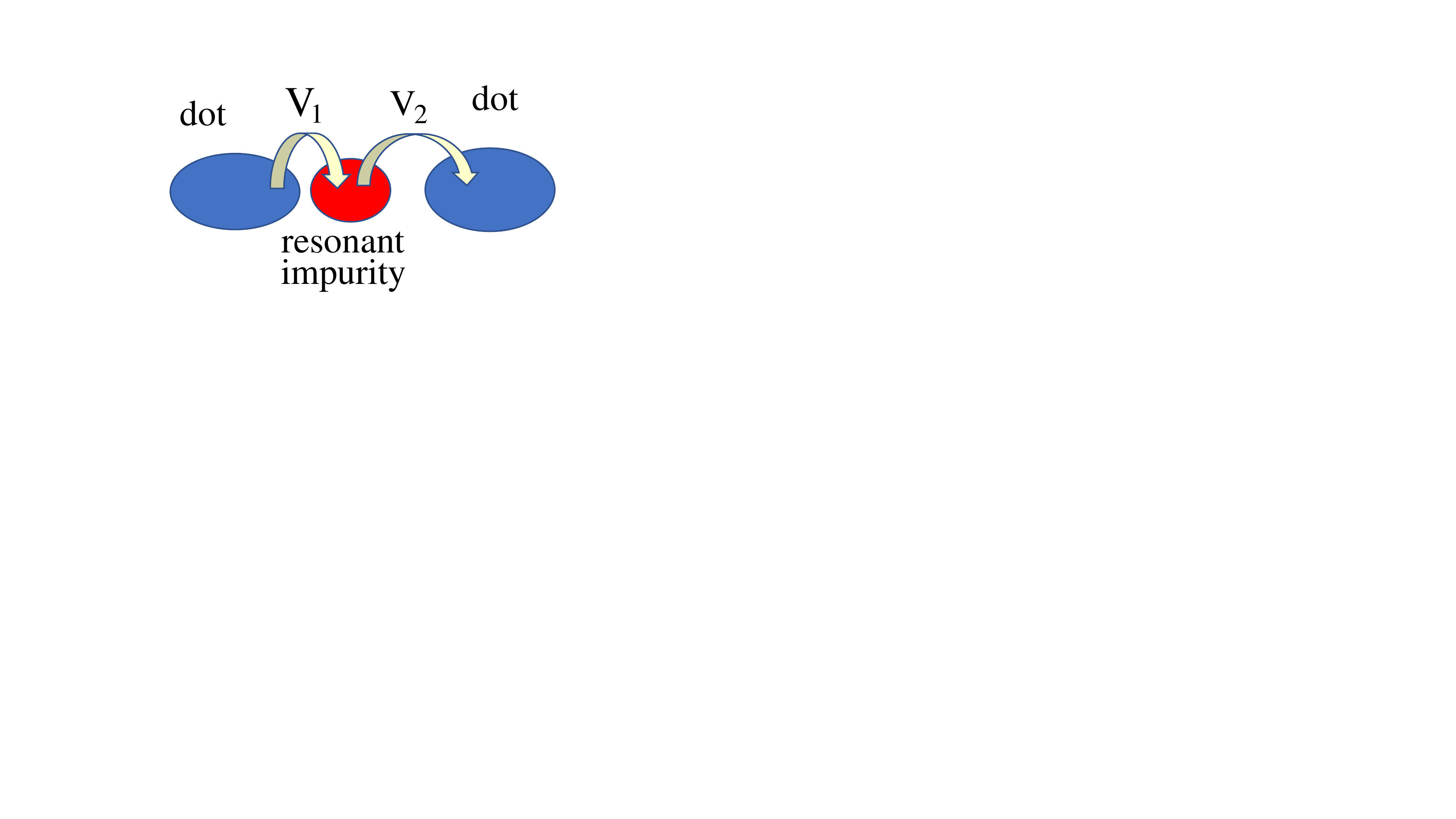}
\caption{(Color online) Illustration of the resonant tunneling 
 configuration. Impurity level  is coupled to the level in the left dot 
 with a coupling coefficient $V_1$ and to the level in the right dot with 
  a coupling coefficient $V_2$. }
\end{figure}

\section{Resonant tunneling current}

Resonant-tunneling configuration is depicted in Fig. 1.
Let $a_1$ and  $a_2$ be the amplitudes to find electron in the left and right dots, respectively, while $c_0$ is the amplitude to find an electron on the impurity. These amplitudes are related by the system of equations
\begin{align}
\label{basic}
&i{\dot a}_1=\frac{vt}{2}a_1+V_1c_0, \\
&i{\dot a}_2=-\frac{vt}{2}a_2+V_2c_0,\\
&i{\dot c}_0={\varepsilon}_{\s 0}c_0+V_1a_1+V_2a_2.
\end{align}
 Here $v$ is the velocity with which the levels in the dots are swept past each other. The impurity level,   $\varepsilon_{\s 0}$, is coupled to the dots with coupling coefficients  
 $V_1$ and $V_2$.  As the levels are swept, they cross pairwise at time moments 
 \begin{equation}
 \label{moments}
 t_1=\frac{2\varepsilon_{\s 0}}{v},~~~~t_2=-\frac{2\varepsilon_{\s 0}}{v},~~~~t_{12}=0.
 \end{equation}Individual Landau-Zener transitions take place at $t=t_1$ and at $t=t_2$  (see Fig. 2), 
 while at $t=t_{12}$ the transition does not occur, since the dot levels are decoupled. 
 
 If individual transitions take place {\em independently}, their respective probabilities are given by the Landau-Zener formula 
 \begin{equation}
 \label{probability}
 {\cal P}_{1}=1-\exp\Bigg(-\frac{2\pi V_1^2}{\frac{1}{4}v}\Bigg),~~~~ {\cal P}_{2}=1-\exp\Bigg(-\frac{2\pi V_2^2}{\frac{1}{4}v}\Bigg).
 \end{equation}
 Velocity $\frac{v}{4}$ in the denominator of the exponents is the arithmetic average of the velocity $\frac{v}{2}$ of the dot level and the velocity of the impurity level, which is zero.

 Another consequence of independence of the partial transitions is that the net probability, $ {\cal P}$,  of tunneling between the dots is equal to the product  ${\cal P}_1{\cal P}_2$.  Our goal is to calculate ${\cal P}$ in the domains when the above decoupling is not permitted. Then the resonant current is expressed via ${\cal P}$ as  
 
 \begin{equation}
 I=\frac{e}{\tau}{\cal P},
 \end{equation}
 where $\tau$ is the period of the voltage sweep.
 
 \begin{figure}
\label{FF}
	\includegraphics[scale=0.74]{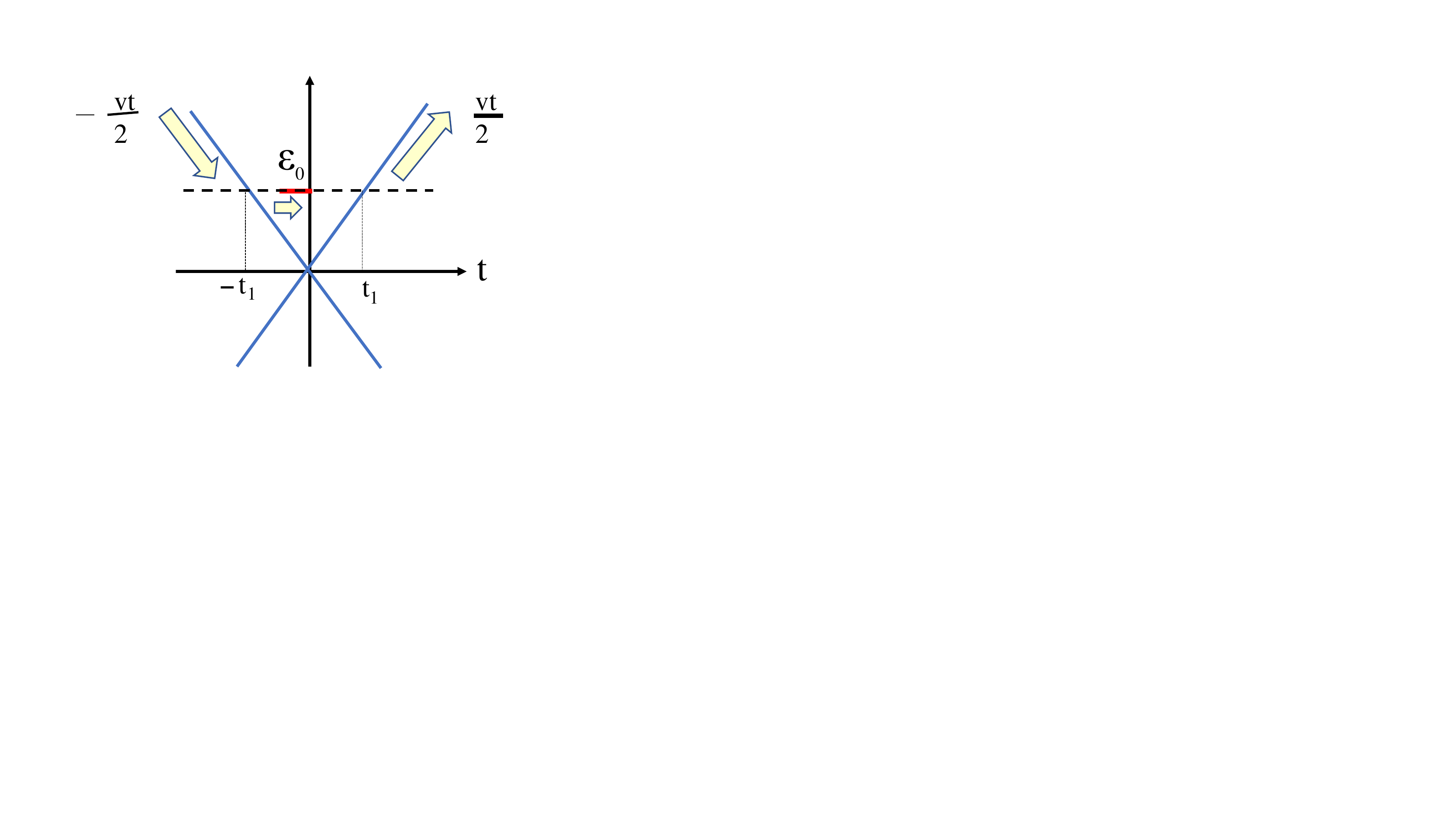}
	\caption{(Color online) Levels in the left and the right dot are swept 
  by the gate voltage as $\frac{vt}{2}$ and $-\frac{vt}{2}$, respectively.
  At time moment $t_1=\frac{2\varepsilon_{\s 0}}{v}$ resonant level crosses the level
  in the left dot, while at time moment $t_2=-\frac{2\varepsilon_{\s 0}}{v}$ it crosses the level in the right dot.  As a result of two Landau-Zener transitions the electron is transferred between the dots.  }
\end{figure}
 
 \section{Semiclassical treatment}

Assuming that   $a_1$, $a_2$ and $c_0$ are proportional to the factor
$\exp\left[ i\int\limits_{\cal C} ^t dt' {\cal L}(t')    \right]$ , while the prefactors are the slow functions of time.   Substituting this form into the system  Eq. \ref{basic}  we  obtain 
\begin{align}
&a_1=-\frac{V_1c_{\s 0}}{{\cal L}+\frac{vt}{2}},~~
a_2=-\frac{V_2c_{\s 0}}{{\cal L}-\frac{vt}{2}}, \label{a}\\
&c_{\s 0}=-\frac{V_1a_1+V_2a_2}{{\cal L}+\varepsilon_{\s 0}}.\label{c}
\end{align}
Combining  Eqs. \ref{a} and \ref{c} leads to the following cubic equation for ${\cal L}$
\begin{align}
\label{cubic}
&\left[ {\cal L}^2-\left( \frac{vt}{2}   \right)^2 \right]\left(  {\cal L}+\varepsilon_{\s 0}    \right)\nonumber\\
&=\Big( V_1^2+V_2^2      \Big ) {\cal L}+\Big(V_2^2-V_1^2\Big)\frac{vt}{2}.
\end{align}
We will analyze how the  solutions of this equation evolve upon gradual decrease of $\varepsilon_{\s 0}$.
 
\subsection{$\large{\varepsilon}_{\s 0}$}

In the lowest order in couplings $V_1$ and $V_2$ the roots of Eq.  \ref{cubic}  are ${\cal L}_{\s 0}=-\varepsilon_{\s 0}$   and ${\cal L}_{\pm}=\pm \Big( \frac{vt}{2}  \Big)$.  The first root corresponds to  the resonant tunneling  of electron via the impurity, while the other two roots describe the conventional tunneling. To find the coupling-dependent correction to the first root we set ${\cal L}_{\s 0}=-\varepsilon_{\s 0}$
in the square brackets and in the right-hand side of Eq. \ref{cubic}. This yields

\begin{equation}
\label{root}
{\cal L}_{\s 0}=-\varepsilon_{\s 0}+\frac{-\varepsilon_{\s 0}(V_1^2+V_2^2)+\frac{vt}{2}(V_2^2-V_1^2)}{\varepsilon_{\s 0}^2-\left(\frac{vt}{2}\right)^2}.
\end{equation}
The assumption that the second term in Eq. \ref{root} is a correction to $\varepsilon_{\s 0}$ amounts to the condition $(V_1^2+V_2^2) \ll \varepsilon_{\s 0}^2$,  i. e.  the condition that the impurity level is high enough.

The other two roots of  Eq. \ref{cubic} are obtained by neglecting ${\cal L}$ in the bracket $\big( {\cal L}+\varepsilon_{\s 0}  \big)$. Then the equation  Eq.  \ref{cubic} becomes quadratic and  can be cast in the form
\begin{equation}
\label{quadratic}
\left[ {\cal L}_{\pm}-\frac{V_2^2+V_1^2}{2\varepsilon_{\s 0}} \right]^2=
\left[\frac{vt}{2}+\frac{V_2^2-V_1^2}{2\varepsilon_{\s 0}} \right]^2+\left(\frac{V_1V_2} {\varepsilon_{\s 0}}   \right)^2.
\end{equation}
Upon an appropriate  shifts of  ${\cal L}$ and $t$, Eq. (\ref{quadratic})  can be reduced to a standard 
form,  ${\cal {\tilde L}}^2=\Big( \frac{v{\tilde t}}{2}\Big)^2 +J^2$,  corresponding to the gap $J=\frac{V_1V_2}{\varepsilon_{\s 0}}$.     The  probability of  the electron  transfer is expressed  via this gap in the  form conventional for the Landau-Zener transition
\begin{align}
\label{high}
&{\cal P}=
1-\exp\left(-\frac{2\pi J^2}{v}  \right)\nonumber\\
&=1-\exp\left[-\frac{2\pi}{v}  \Bigg(\frac{V_1V_2}{\varepsilon_{\s 0}} \Bigg)^2     \right].
\end{align}
Naturally,  if any of two couplings is zero, there is no electron transfer.  It is less trivial that, the higher is the impurity level, the smaller is the transfer probability.  Note that, within the
factor of the density of states square, the effective tunnel parameter $J^2$ in the exponent can be identified with   the cotunneling conductance.  Recall that the result Eq.  \ref{high} was obtained in the semiclassical approximation, i.e. under the condition that the exponent in  
Eq. \ref{high} is big.  It is known that, for  conventional Landau-Zener transition, the exact result coincides with semiclassical.  Concerning the result Eq. \ref{high}, it emerges without semiclassics if we neglect ${\dot c}$ in Eq. \ref{basic} and introduce  instead of $a_1$ and $a_1$ the new variables

\begin{align}
\label{tilde}
&{\tilde a}_1=a_1\exp\left[{-i\left(\frac{V_1^2+V_2^2}{2\varepsilon_{\s 0}}\right)t}\right], \nonumber\\
&{\tilde a}_2= a_2\exp\left[-{i\left(\frac{V_1^2+V_2^2}{2\varepsilon_{\s 0}}\right)t}\right].
\end{align}
With these new variables,  the system \ref{basic} takes the form
\begin{align}
\label{basic2}
&i{\dot {\tilde a}}_1= \left(\frac{vt}{2}-\frac{V_1^2-V_2^2}{2\varepsilon_{\s 0}}\right) {\tilde a}_1        -\frac{V_1V_2}{\varepsilon_{\s 0}}{\tilde a}_2, \\
&i{\dot {\tilde a}}_2= -\left(\frac{vt}{2}-\frac{V_1^2-V_2^2}{2\varepsilon_{\s 0}}\right)       {\tilde a}_2-\frac{V_1V_2}{\varepsilon_{\s 0}}{\tilde a}_1,
\end{align}
for which the result Eq. \ref{high} is exact. The condition that the gap is much smaller than
$\varepsilon_{\s 0}$ ensures that ${\dot c}_0$ in Eq. \ref{basic} can be neglected. 

It is instructive to compare the probabilities of transfer in the cotunneling regime  and in the independent crossing regime  when these probabilities are small
\begin{equation}
\label{perturbatiive}
{\cal P}\Big\vert_{\text{cotunnel}}\approx \frac{2\pi}{v}\left( \frac{V_1V_2}{\varepsilon_{\s 0} } \right)^2,~ {\cal P}\Big\vert_{\text{independent}}\approx \left( \frac{8\pi V_1V_2} {v}  \right)^2.
\end{equation}
It follows from Eq. \ref{perturbatiive} that when the couplings $V_1$, $V_2$ are small the criterion for cotunneling regime to apply is $\varepsilon_{\s 0}^2\gg v$.

\begin{figure}
	\label{FFF}
	\includegraphics[scale=0.58]{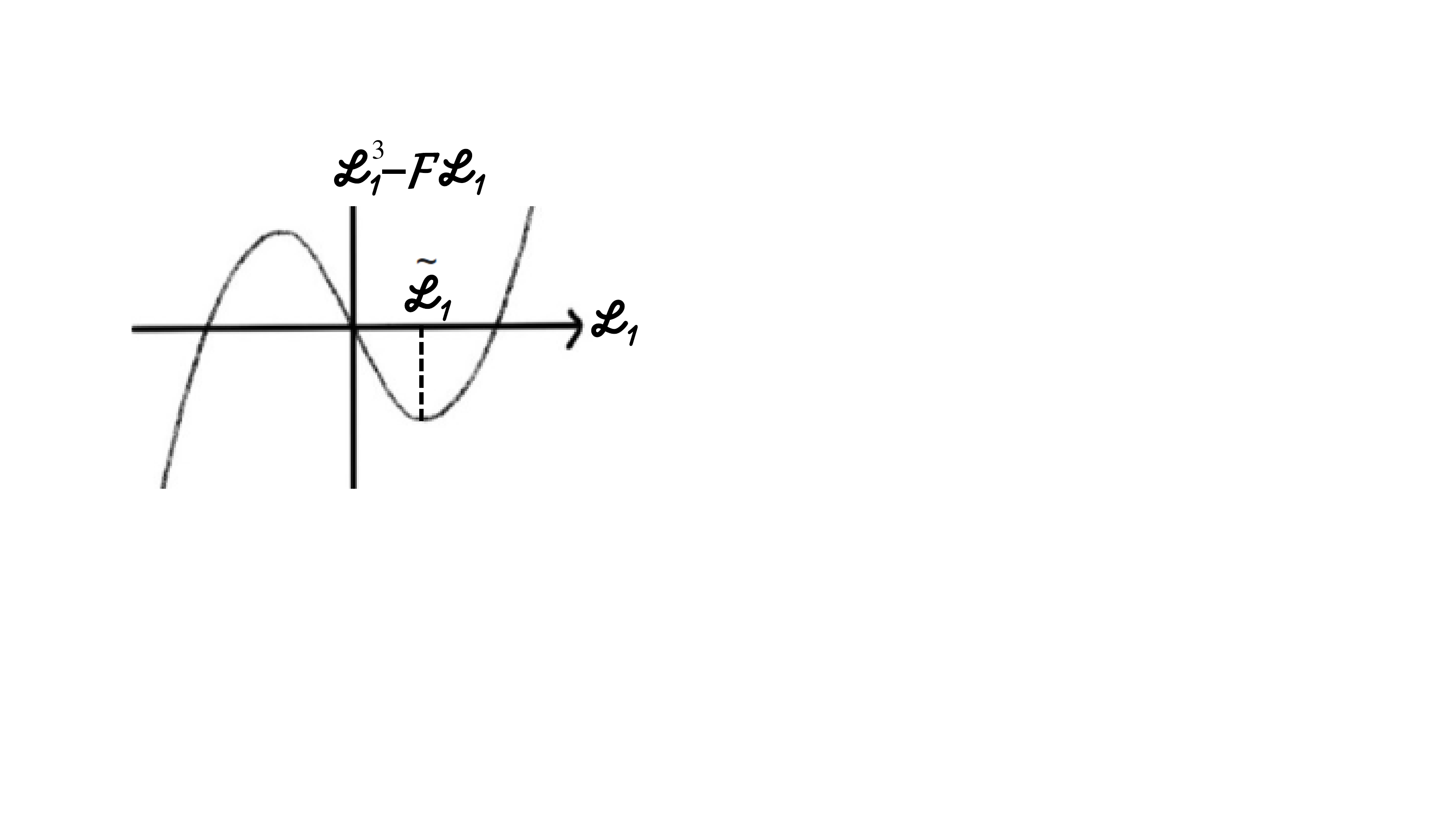}
\caption{(Color online) Graphic illustration of the solution of Eq.  \ref{CUBIC}. In the vicinity of the minimum at ${\cal L}_1=\tilde{\cal L}_1$ two roots of the cubic equation are close to each other. Then the transport between the dots is described by the conventional Landau-Zener theory. }
\end{figure}

\subsection{Upon decreasing ${\varepsilon}_{\s 0}$}
To analyze the cubic equation Eq. \ref{cubic} in this limit we eliminate the quadratic term by shifting the variable ${\cal L}$ as follows

\begin{equation}
\label{substitution}
{\cal L}=-\frac{\varepsilon_{\s 0}}{3}+{\cal L}_1.
\end{equation}Then the exact equation for ${\cal L}_1$ takes the form 
\begin{equation}
\label{CUBIC}
{\cal L}_1^3-{\cal L}_1F=G,
\end{equation}
where the functions $F$ and $G$ are defined as
\begin{align}
\label{FG}
&F=\frac{\varepsilon_{\s 0}^2}{9}+\left(\frac{vt}{2}\right)^2+V_1^2+V_2^2, \nonumber\\
&G=-\frac{\varepsilon_{\s 0}}{3}\left[\frac{2\varepsilon_{\s 0}^2}{9}-2\left(\frac{vt}{2}\right)^2+V_1^2+V_2^2\right]+\frac{vt}{2}\Big(V_2^2-V_1^2\Big).
\end{align}
The left-hand side of Eq. \ref{CUBIC} has a minimum at ${\cal L}_1 ={\tilde{\cal L}_1}=\left(\frac{F}{3}  \right)^{1/2}$, see Fig. 3.  
The minimal value is equal to
$-2\left(\frac{F}{3}\right)^{3/2}$.  Consider a particular case  $V_2=V_1=V$ and $t=0$. Then Eq. \ref{CUBIC} takes the form
\begin{equation}'
\label{CUBIC1}
\left[1+\frac{1}{18}\left( \frac{\varepsilon_{\s 0}}{V}\right)^2\right]^{3/2}=\left( \frac{3}{8} \right)^{1/2}\left(\frac{\varepsilon_{\s 0}}{V} \right)\left[1+\frac{1}{9}\left(\frac{\varepsilon_{\s 0}}{V} \right)^2  \right].
\end{equation}
The solution of the cubic equation is $\varepsilon_{\s 0}=\tilde{ \varepsilon_{\s 0}   }=1.55V$.

To find the time dependence of ${\cal L}_1$, we restore the time dependence in the functions $F$ and $G$ and expand 
${\cal L}_1^3-{\cal L}_1F$ in Eq. \ref{CUBIC} near the minimum. Then the time-dependent correction to the solution of Eq. \ref{CUBIC1} reads

\begin{equation}
\label{expanded}
\left[ {\cal L}_1-\left( \frac{F}{3}    \right) ^{1/2}     \right]^2=\frac{G+2\left( \frac{F} {3}  \right)^{3/2}}
{3 \left( \frac{F} {3}  \right)^{1/2}       }.
\end{equation}
At $t=0$ the right-hand side of Eq. \ref{expanded} turns to zero. Upon expanding $G$ and $F$ around $\tilde{\varepsilon}_{\s 0}$ and restoring  $\left(\frac{vt}{2}\right)^2$ in $G$ and $F$, Eq. \ref{expanded} takes the form

\begin{equation}
\label{expanded1}
\left[ {\cal L}_1-\left( \frac{F}{3}    \right) ^{1/2}     \right]^2=\alpha\left(\frac{vt}{2}\right) ^2
+\beta\Big(\varepsilon_{\s 0} -\tilde{\varepsilon_{\s 0}}\Big),
\end{equation} where the coefficients of expansion $\alpha$ and $\beta$ are of the order of $1$. We see that, for $\varepsilon_{\s 0}>\tilde{\varepsilon}_{\s 0}$, two  of the three solutions of the cubic equation Eq. \ref{cubic} satisfy the quadratic equation having the Landau-Zener form Eq. \ref{quadratic}.

Comparing Eq. \ref{expanded1} to Eq. \ref{high} leads immediately to the expression for the transition probability

\begin{equation}
\label{threshold}
{\cal P}=
1-\exp\left[-\frac{2\pi}{v} \left(\frac{\beta}{\alpha^{1/2}}  \right)\Big(\varepsilon_{\s 0}-
\tilde{\varepsilon_{\s 0}}\Big)\right].
\end{equation}

To summarize our semiclassical analysis of the   transmission probability, ${\cal P}$, for a symmetric position of the resonant level, $V_1=V_2=V$, it depends on the dimensionless ratio,  $\Big\vert\frac{V}{\varepsilon_{\s 0}}     \Big\vert$.   When this ratio is small, the transport between the dots is due to cotunneling, and ${\cal P}$ is small. As $\varepsilon_{\s 0}$ is gradually shifted down, ${\cal P}$ increases and passes through a maximum at ${\varepsilon}_0 \sim V$.  All three states,  left dot, impurity level, and the right dot are represented in the wave-function  approximately equally. As the level reaches a critical  value $\varepsilon_{\s 0}  =\tilde{\varepsilon}_{\s 0}=1.55V$ from the above, the transmission probability is small again, as follows from Eq. \ref{threshold}. Upon further decrease of $\varepsilon_{\s 0}$, the transmission probability keeps dropping down to the value $\left( {\cal P}  \right)^2$, predicted by the independent crossing approximation.

\subsection{Perturbative treatment}

Perturbative  treatment applies when the couplings of the impurity to the dots differ strongly.  Then the picture of  the electron transfer between the dots gets strongly simplified.  Assume e.g. that $V_2\ll V_1$. This relation implies that beatings of electron 
between the left dot and impurity is much faster than the beatings between the impurity
and the right dot. In this  limiting case,  the Landau-Zener transition between the left dot and impurity gets {\em completed} before electron accomplishes a {\em single} virtual tunneling from the impurity to the right dot. To quantify this scenario,  the Landau-Zener time  can be estimated as  $\frac{V_1}{v}$,   while the beating time can be estimated as $\frac{1}{V_2}$.  Then the condition that the second time is longer reads $V_1V_2 \ll v$. Under this condition, the probability of  the transfer of the electron from the impurity to the second dot can be calculated perturbatively.

At time $t=t_1=2\frac{\varepsilon_{\s 0}}{v}$ the dot level crosses the impurity level.  To capture the vicinity of this crossing we introduce a shifted time
\begin{equation}
\label{new}
T=t-t_1
\end{equation}
and the new variables
\begin{align}
\label{variables}
&a_1=A_1(T)\exp\left( -i\varepsilon_{\s 0}T-\frac{vT^2}{8} \right),\nonumber\\ 
&a_2=A_2(T)\exp\left( -i\varepsilon_{\s 0}T-\frac{vT^2}{8} \right),\nonumber\\
&c_0=C_0(T)\exp\left( -i\varepsilon_{\s 0}T-\frac{vT^2}{8} \right).
\end{align}
With these variables the system Eq. \ref{basic} acquires the form 

\begin{align}
\label{basic1}
&i{\dot A}_1=\frac{vT}{4}A_1+V_1C_0, \\
&i{\dot A}_2=-\frac{v}{2}\left(\frac{3T}{2}+\frac{4\varepsilon_{\s 0}}{v}\right)A_2+V_2C_0,\\
&i{\dot C}_0=-\frac{vT}{4}C_0+V_1A_1+V_2A_2.
\end{align}
If we neglect $V_2$, the system gets decoupled. Then the first and the third equations reduce to the system describing a conventional Landau-Zener transition with matrix element $V_1$ and velocity $\frac{v}{4}$. A textbook expression for $A_1$ and  $C_0$, which satisfies the initial condition $C_0(-\infty) =0$, reads

\begin{equation}
\label{C0}
A_1(T)=D_{\nu}\Big({\sqrt \frac{v}{4}}e^{\frac{\pi i}{4}} T    \Big),~               C_0(T)=-i{\sqrt \nu}D_{\nu-1}\Big({\sqrt \frac{v}{4}}e^{\frac{\pi i}{4}} T    \Big),
\end{equation}
where $D_{\nu}(z)$  and $D_{\nu-1}(z)$  are the parabolic cylinder functions, whereas the parameter $\nu$ is defined as
\begin{equation}
\label{nu}
\nu=-i\frac{V_1^2}{\Big(\frac{v}{4}\Big)}.
\end{equation}

Perturbative incorporation of $V_2$ reduces to solving the second equation of the system for $A_2$ while assuming that $C_0(T)$ is fixed. Upon imposing a natural initial condition $A_2(-\infty) =0$, we get

\begin{align}
\label{perturbative}
&{\Big\vert} A_2(T) {\Big\vert}^2 \nonumber\\
&=
V_2^2 {\Bigg\vert} \int\limits_{-\infty}^T dT'  C_0(T')
\exp\left[-\frac{iv}{2}\Bigg( \frac{3T'^2}{4}+\frac{4\varepsilon_{\s 0}T'}{v}         \Bigg)      \right]{\Bigg\vert}^2.  
\end{align}
Substituting $C_0(T)$ from Eq. \ref{C0} and introducing a new variable 
$z=T\sqrt{\frac{v}{4}}\exp\Big(\frac{\pi i}{4}\Big)$,  we find the following expression for $\vert A_2(\infty)\vert^2$

\begin{align}
\label{A2}
&\vert A_2(\infty) \vert^2
&=\frac{4|\nu|V_2^2}{v}\Bigg\vert \int\limits_{-\infty} ^{\infty}dz             D_{\nu-1}(z)\exp\Big[-\frac{3z^2}{2}-\kappa z         \Big]\Bigg\vert^2,            
\end{align}
where the parameter $\kappa$ is defined as
\begin{equation}
\label{kappa}
\kappa=\frac{4\varepsilon_{\s 0}}{v^{1/2}}\exp\Big(\frac{\pi i}{4}\Big).
\end{equation}
To evaluate the integral in Eq.  \ref{A2} we use the following representation\cite{Representation} of the parabolic cylinder function

\begin{equation}
	\label{representation}	
	D_{\nu-1}(z)=-\Bigl(\frac{2}{\pi} \Bigr)^{1/2}e^{\frac{z^2}{4}}\int\limits_0^{\infty}	du~
	e^{-\frac{u^2}{2}}u^{\nu}\sin\left(zu-
	\frac{\pi\nu}{2}  \right).
\end{equation}
Upon substituting this representation into Eq.  \ref{A2}, the integral over $z$ becomes gaussian leading to 
 
\begin{align}
\label{A22}
&\vert A_2(\infty) \vert^2\nonumber\\
&=\frac{16|\nu|V_2^2}{3v}\Bigg\vert \int\limits_{0} ^{\infty} du u^{\nu}\exp\Big(-\frac{2}{3}u^2   \Big)  \sin\Big( \frac{\kappa u}{3}+\frac{\pi \nu}{2}         \Big)\Bigg\vert^2.
\end{align}
Comparing Eqs. \ref{representation} and \ref{A22},  we conclude that the integral $du$ can  also be expressed via the function $D_{\nu-1}$

\begin{equation}
\label{A222}
\vert A_2(\infty) \vert^2=\frac{2\pi |\nu|V_2^2}{v}\Bigg\vert  D_{\nu-1}      \Big(   -\frac{\kappa}{2\sqrt{3}}\Big)\Bigg\vert ^2.
\end{equation}Note now, that the combination $\frac{|\nu|V_2^2}{v}$ in the prefactor with definition Eq. (\ref{nu})  of the parameter $\nu$ assumes the form $\Big(\frac{V_1V_2} {v}  \Big)^2$,  which is the square of the parameter established above from the qualitative consideration. Smallness of this parameter ensures the applicability of the perturbative treatment.

The factor $\Big\vert  D_{\nu-1}      \Big(   -\frac{\kappa}{2\sqrt{3}}\Big)\Big\vert ^2$ describes , via the argument $\kappa$,  the dependence of the tunnel conductance on the level position $\varepsilon_{\s 0}$.  At resonance, $\varepsilon_{\s 0}=0$, using the representation Eq.   \ref {representation}, one finds $\Big\vert  D_{\nu-1}      \big(  0\big)\Big\vert ^2=
\frac{\sinh^2\frac{\pi |\nu|}{2}}{  \cosh\frac{\pi |\nu|}{2}}$.
Correction to the conductance at finite $\varepsilon_0$ follows from the expansion
\begin{align}
\label{RATIO}
&\Bigg\vert  \frac{   D_{\nu-1}      \Big(   -\frac{\kappa}{2\sqrt{3}}\Big)    }  {D_{\nu-1}(0)}    \Bigg\vert^2\nonumber\\
&\approx 1 -\left( \frac{32}{3}  \right)^{1/2}\frac{\varepsilon_{\s 0}}{v^{1/2}} 
\text{Re}\Bigg\{ \frac{\Gamma(\frac{\nu}{2}+1)}     {\Gamma(\frac{\nu}{2} +\frac{1}{2}) }e^{
\frac{\pi i}{4}}\Bigg\}.
\end{align}

As was discussed above, see Eq. \ref{perturbatiive},  it is the   parameter $\frac{\varepsilon_{\s 0}^2}{v}$ that should be large  for cotunneling regime to unfold. 

Note, that there is a formal correspondence between the standard Landau-Zener result Eq. \ref{C0} for $C_0(T)$ and our perturbative result \ref{A222} for $A_2(\infty)$. It is well-known, see e.g. 
Refs.   \onlinecite{review1,review2}, that the Landau-Zener transition  is accompanied by the oscillations of the level occupation {\em with time}. Then we conclude that  $|A_2(\infty)|^2$ contains an oscillating term  {\em as a function of the resonant level position}.  Building up on the formal correspondence between $|C_0(T)|^2$ and $|A_2(\infty)|^2$, we can readily write down the expression for this oscillating term 

\begin{equation}
\label{oscillating}
\cos\Bigg[\frac{2\varepsilon_{\s 0}}{v^{1/2}} \left( |\nu|+\frac{4\varepsilon_{\s 0}^2}{v} \right)^{1/2} -\Psi  \Bigg],
\end{equation}
where the phase $\Psi$ is given by

\begin{equation}
\label{PHASE}
\Psi=|\nu|\ln \Bigg\{\frac{ \left( |\nu| +
\frac{4\varepsilon_{\s 0}^2}{v}\right)^{1/2}-\frac{2\varepsilon_{\s 0}}{v^{1/2}}       }
{  \left( |\nu| +\frac{4\varepsilon_{\s 0}^2}{v}\right)^{1/2}+\frac{2\varepsilon_{\s 0}}{v^{1/2}}        }\Bigg\}.
\end{equation}
While the term Eq. \ref {oscillating} is a small correction to the monotonic term, the oscillations of the resonant-tunneling current with the level position are highly nontrivial.

\section{Comparison to three-level  bow-tie model}

The paper Ref. \onlinecite{Carroll1} is entirely devoted to the transfer probability in the three-level model in the limit $\varepsilon_{\s 0}\rightarrow 0$. Quantitatively,  this limit corresponds to  $|\varepsilon_{\s 0}|\ll V_1, V_2$.   Similar to the present paper, it was assumed in Ref.   \onlinecite{Carroll1}  that the first and the third levels are completely decoupled. 
Since in the limit $|\varepsilon_{\s 0}|\ll V_1, V_2$,  the position  ${\varepsilon}_{\s 0}$  should drop out from the transfer probability, it can depend  only on  two dimensionless combinations: $\frac{V_1^2}{v}$ and  $\frac{V_2^2}{v}$.  Using the contour integral representation,\cite{Demkov1968}   it was shown in Ref.  \onlinecite{Carroll1}, see also Ref. \onlinecite{Carroll}, that the result ${\cal P}={\cal P}_1{\cal P}_2$,  corresponding to the independent crossing, is valid. Thus, upon increasing of $\varepsilon_{\s 0}$ ,   the probability, ${\cal P}$, crosses over from independent crossing to the cotunneling regime Eq. \ref{high}, which applies for  $\varepsilon_{\s 0}^2\gg (V_1^2+V_2^2). $       The result Eq. \ref{high} seems to be the only  result predicting  the {\em analytical} dependence of ${\cal P}$ on the level position, $\varepsilon_{\s 0}$.  This is certainly in agreement with conclusion of  Ref.~\onlinecite{dot1},  supported numerically,  that the independent crossing approximation fails upon increasing of  $\varepsilon_{\s 0}$.  Equally, the result  Eq. \ref{A222}  of the perturbative  treatment   indicates the  explicit dependence of the tunnel probability on $\varepsilon_{\s 0}$.  

General approach in the papers \cite{Carroll,Carroll1,Bow1,Bow2,Bow3,Bow5} is based on the representation of the solution of the system Eq. \ref{basic} in the form of the contour integral. The approach
adopted in the present paper is based on exploiting of the small parameter in the problem. In the three-state model, the semiclassical equation for the quasienergy, ${\cal L}$, has three solutions. Small parameter emerges 
when one of the solutions is much  bigger than the other two.  Neglecting the fast solution and keepinng only two slow solutions\cite{Rajesh0,Rajesh1} allows to reduce the problem to the effective Landau-Zener transition.  Other small parameter  emerges when the coupling of impurity level  to one dot is much bigger than the coupling  to the other dot.

\end{document}